\documentclass{article}
\usepackage{spconf,amsmath,graphicx,adjustbox,hyperref}
\usepackage{color}
\usepackage{siunitx}
\usepackage{booktabs}
\usepackage{caption}
\usepackage{multirow}
\usepackage{cite}

\title{AutoPrep: An Automatic Preprocessing Framework for In-the-Wild Speech Data}
\name{Jianwei Yu$^{1}$, Hangting Chen$^{1}$, Yanyao Bian$^{1}$, Xiang Li$^{1}$, Yi Luo$^{1}$, Jinchuan Tian$^{1}$, \\  \textit{Mengyang Liu$^{1}$}, \textit{Jiayi Jiang$^{1}$}, \textit{Shuai Wang$^{2}$}}
\address{$^{1}$Tencent AI Lab, $^{2}$Shenzhen Research Institute of Big data}
\begin{document}
\ninept
\maketitle
\begin{abstract}
Recently, the utilization of extensive open-sourced text data has significantly advanced the performance of text-based large language models (LLMs). 
However, the use of in-the-wild large-scale speech data in the speech technology community remains constrained. 
One reason for this limitation is that a considerable amount of the publicly available speech data is compromised by background noise, speech overlapping, lack of speech segmentation information, missing speaker labels, and incomplete transcriptions, which can largely hinder their usefulness. On the other hand, human annotation of speech data is both time-consuming and costly.
To address this issue, we introduce an automatic in-the-wild speech data preprocessing framework (AutoPrep) in this paper, which is designed to enhance speech quality, generate speaker labels, and produce transcriptions automatically.
The proposed AutoPrep framework comprises six components: speech enhancement, speech segmentation, speaker clustering, target speech extraction, quality filtering and automatic speech recognition.
Experiments conducted on the open-sourced WenetSpeech and our self-collected AutoPrepWild corpora demonstrate that the proposed AutoPrep framework can generate preprocessed data with similar DNSMOS and PDNSMOS scores compared to several open-sourced TTS datasets. 
The corresponding TTS system can achieve up to 0.68 in-domain speaker similarity.\footnote{Demos and more details can be found in \url{https://tomasjwyu.github.io/AutoPrepDemo/}} 
\end{abstract}
\begin{keywords}
Speeech data preprocessing, speech enhancement, speaker clustering
\end{keywords}

\vspace{-0.3cm}
\section{Introduction}
\label{sec:intro}

Gathering large-scale, high-quality training data with comprehensive and accurate labels has always been a critical aspect of speech technology development. 
Over the past few decades, the speech community has devoted considerable time and effort to manually recording, collecting, and annotating a vast amount of speech data with corresponding segmentation, transcription and speaker labels, which has significantly advanced the performance of various speech technologies, such as automatic speech recognition (ASR) \cite{asr_survey, radford2023robust, zhang2023google}, text-to-speech synthesis (TTS) \cite{tts_survey, kharitonov2023speak, wang2023neural, jiang2023mega}, speaker verification (SV) \cite{sv_survey, caron2021emerging}, and speech enhancement (SE) \cite{se_survey}. 
However, the volume of accessible human-annotated speech data is still limited for further improving the performance and generalization of current speech models, especially for TTS which requires high-quality speech recordings with multiple speakers and styles. 
To this end, leveraging the vast amount of open, in-the-wild speech data from open video recordings, podcasts and audiobooks presents a promising approach.

However, directly utilizing in-the-wild speech data poses challenges in two main aspects:
First, the absence of necessary annotations, such as reliable text transcriptions, segmentation information, and speaker labels, impedes the direct use of this data for supervised training tasks, such as ASR, TTS and SV. 
Second, simply adding more data does not guarantee performance enhancement. Low-quality data with unexpected background noise, reverberation, speech overlap, and distortion, can significantly degrade the performance of generation tasks like TTS.
For the missing annotation issue, previous research has focused primarily on unsupervised pre-training methods \cite{wav2vec, baevski2020wav2vec, hubert, wavlm, borsos2023audiolm}. 
However, for most downstream tasks, these methods still require annotated speech data for fine-tuning the pretrained models. 
Regarding the second issue, prior research \cite{koizumi2023libritts, valentiniinvestigating} has proposed leveraging speech enhancement methods to handle unstable speech quality in TTS model construction. 
However, these methods are only effective in part of the scenarios presented in diverse in-the-wild data. 
Nevertheless, to the best of our knowledge, there is limited previous work that attempts to comprehensively address both challenges associated with directly using in-the-wild speech data.

Therefore, in this paper, we propose a pipelined automatic in-the-wild speech data preprocessing framework, named AutoPrep, which aims to directly solve the aforementioned two issues by automatically generating segmentation information, speaker labels, and transcriptions for in-the-wild speech data while removing background noise, dereverberation, and speech overlap to improve speech quality. 
To provide high-quality speech and annotation, we incorporate competitive models into the proposed AutoPrep framework, including the Band-Split RNN (BSRNN) speech enhancement model \cite{bsrnn}, WeSpeaker speaker embedding model \cite{wespeaker}, personalized BSRNN (pBSRNN) target speech extraction (TSE) model \cite{yu2023tspeech}, and a 60k-hour trained multilingual conformer-based \cite{conformer} RNN-Transducer \cite{rnnt} ASR system. In addition, the DNSMOS \cite{dnsmos} and PDNSMOS \cite{dubey2023icassp} metrics are also employed to evaluate the quality of the processed speech data, filtering out low-quality instances to ensure the speech quality of the final output.
By utilizing the AutoPrep framework, the processed in-the-wild speech data can be directly employed in various tasks, such as TTS, SV, and ASR model training. Users can easily access individual data samples, and selectively choose the desired portions of the data for customized usage with this framework. 
\begin{figure*}[!htb]
	\centering
	\includegraphics[width=1\linewidth]{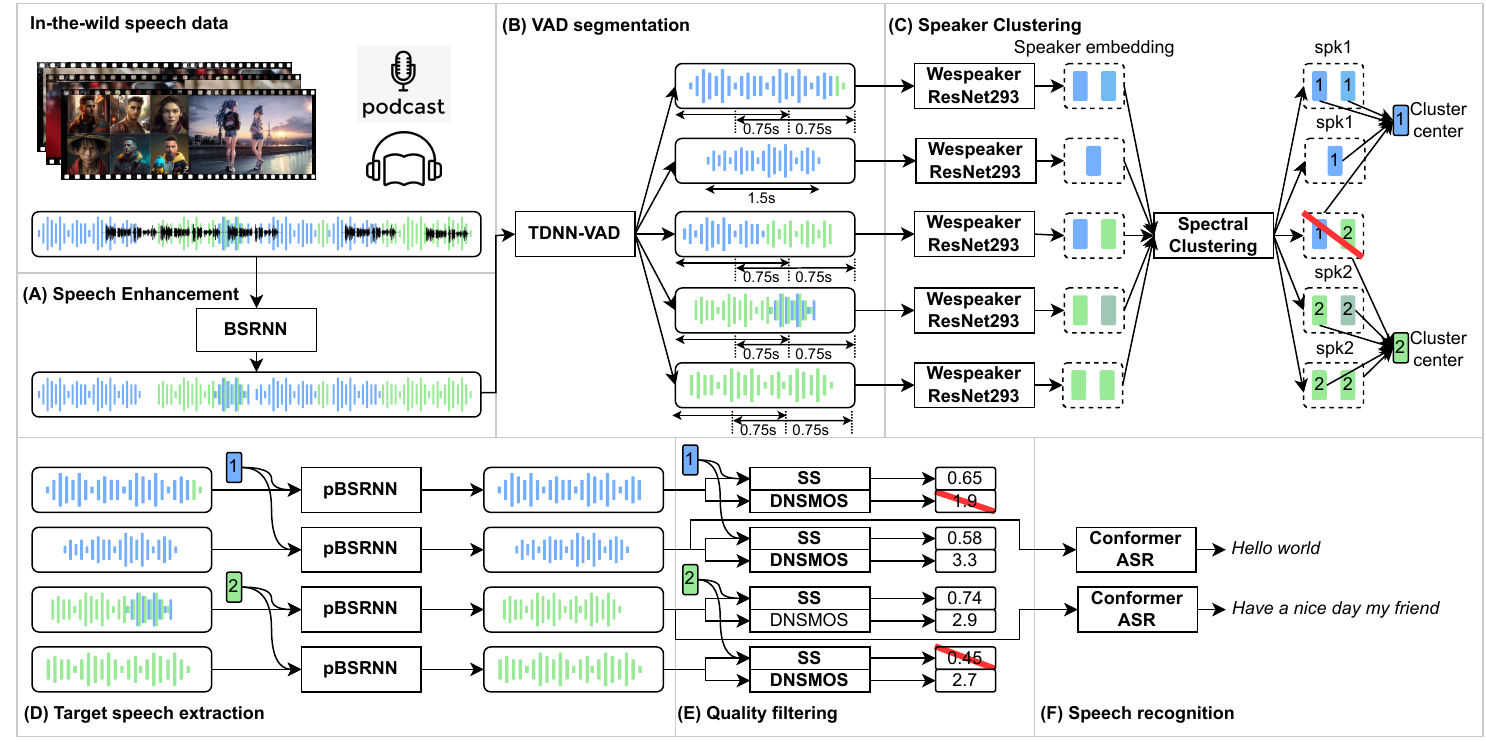}
    \caption{The diagram of the proposed full-band AutoPrep framework. }
\label{fig:AutoPrep}
\end{figure*}

Due to the lack of annotations for in-the-wild data, directly evaluating the annotation performance of the AutoPrep framework can be challenging.
In this paper,  we first apply DNSMOS and PDNSMOS metrics to measure the improvement in speech quality achieved by AutoPrep.
In addition, we use the processed data to train a DurIAN \cite{yu2019durian} multi-speaker TTS system to evaluate the proposed framework.
Specifically, we compute the in-domain speaker similarity (SS) between the reference and synthesized speech to assess the quality of speaker labels and use human MOS scores to overall prepossessing quality.
Experiments conducted on the open-sourced WenetSpeech \cite{zhang2022wenetspeech} and our self-collected AutoPrepWild speech data demonstrate that the proposed AutoPrep framework can generate processed data speech with similar DNSMOS and PDNSMOS scores compared to the open-sourced AIShell-3 \cite{shi2020aishell} and LibriTTS \cite{zen2019libritts} datasets. 
And the corresponding TTS system can achieve up to 0.68 indomain speaker similarity.

The main contribution of this paper can be listed as follows:  1) The proposed AutoPrep framework is among the first works that can automatically improve speech quality and generate comprehensive high-quality annotations for in-the-wild speech data. 2) We conducted an experiment on a TTS system trained using the automatically generated data to validate the effectiveness of the proposed AutoPrep.  3) We open-source the speaker labels of WenetSpeech and GigaSpeech, generated by the proposed AutoPrep framework.\footnote{Processed audio segments of WenetSpeech-P (sub) will be available soon after the final quality check.}

\vspace{-0.3cm}
\section{AutoPrep Framework}
\label{sec:framework}
Figure~\ref{fig:AutoPrep} illustrates the workflow of the proposed AutoPrep framework, where the raw speech data is processed through speech enhancement, voice-active-detection (VAD) based segmentation, speaker clustering, target speech extraction, quality filtering, and speech recognition step-by-step, and ultimately transformed into high-quality annotated speech data. 

\vspace{-0.3cm}
\subsection{Speech enhancement}
To improve the audio quality of the in-the-wild speech data and reduce the impact of noise on the subsequent components in AutoPrep, we first perform speech enhancement on the raw speech. 
In this process, we consider the following aspects:
First, to deal with various sampling rates of the original data, we use a universal-sample-rate USRBSRNN model \cite{yu2023efficient} that supports 8k-48kHz.
In addition, to achieve better speech enhancement performance, we employ a non-streaming BSRNN noise reduction model, which has achieved state-of-the-art (SOTA) results on the DNS2020 test dataset.
Given that the original speech data often exceeds 5 minutes in length, it can be too long for a non-streaming BSRNN model to process directly.  
To address this issue,  we employ a chunk-wise inference approach during the speech-enhancing process. 
Specifically, we divide the raw speech into chunks with a 12s window length and a 4s window shift. 
We then process each chunk with the BSRNN model and extract the middle 4s of the output audio as the chunk-wise output. 
Finally, we concatenate these chunk-wise outputs to produce the enhanced speech data.
We acknowledge that the use of a speech enhancement model can inevitably introduce some unwanted artifacts in the data. 
However, given the strength of the adopted BSRNN model, the benefits it provides to the subsequent tasks such as segmentation, speaker clustering, and ASR can be much more significant than the impact of model distortion. Note that, the speech enhancement component in AutoPrep only deals with background noise, and will keep all speech contents including speech overlapping. 

\vspace{-0.3cm}
\subsection{Speech segmentation}
To exclude non-speech parts from the original data, and obtain effective speech segments, we then use a time-delay neural network (TDNN)-based \cite{tdnn} voice active detection (VAD) model with an active threshold 0.76 to segment the enhanced speech data.
Specifically, to prevent truncating continuous audio, we only split the speech when consecutive silence segments exceed 1s and retain 0.4s of audio before and after each segment.
To avoid having speech segments shorter than 1.5s, we merge shorter segments in the middle of the VAD results until the accumulated audio length exceeds 1.5s before performing the segmentation.
To avoid too long (longer than 30s) speech segments, we split the audio at the first silent frame when the segment length surpasses 30s. Additionally, we directly truncate the audio when the segment length exceeds 40s.

\vspace{-0.3cm}
\subsection{Speaker clustering}
After speech segmentation, we aim to assign speaker labels to each segment. 
To accomplish this, we employ the widely used spectral clustering algorithm\footnote{ Code can be found in \url{https://github.com/wenet-e2e/wespeaker/tree/master/examples/voxconverse/v1}} \cite{wespeaker} to determine both the number of speakers and their assignments for these segments. 
Specifically, we first employ the ResNet293 speaker embedding model from Wespeaker to calculate the speaker embeddings for each segment. 
As illustrated in Figure~\ref{fig:AutoPrep}, we divide each speech segment into chunks using a 1.5s window and a 0.75s shift to compute the speaker embeddings to deal with segments containing more than one speaker. 
Next, we construct a similarity matrix for the computed speaker embeddings based on cosine similarity. 
We then calculate the normalized Laplacian matrix and determine the number of speakers by identifying the index of the largest first-order derivative among the sorted eigenvalues of the Laplacian matrix. 
Subsequently, we apply the standard K-Means algorithm to obtain the cluster centers and assignments. 
To avoid situations where the same person is assigned to different clusters, we further merge the clusters if the cosine similarity between their cluster centers is greater than 0.75.
Finally, we assign the speaker label to each segment based on the cluster assignments of their speaker embeddings. As shown in Figure~\ref{fig:AutoPrep}(C), for segments with speaker embeddings assigned to different clusters, we do not assign any speaker labels to them.
Note that the speaker clustering in AutoPrep is only computed within a limited number of long speech utterances. 
Normally, we restrict the total length of audio processed in a single clustering operation to less than 2 hours.

\vspace{-0.3cm}
\subsection{Target speech extraction}
To address speech overlapping, we further employ a non-streaming personalized BSRNN (pBSRNN) model to extract the target speech for each segment, using the cluster center of each speaker as the enrolled speaker embedding. It is worth noting that this pBSRNN model is a modification of the streaming pBSRNN model we used in the DNS 2023 challenge~\cite{dnsmos}.

\vspace{-0.3cm}
\subsection{Quality filtering}
In real-world scenarios, it is inevitable that some noise cannot be effectively handled by speech enhancement, and clustering algorithms may also have errors. Therefore, to ensure the quality of the final processed speed data, we further employ data filtering methods in AutoPrep. 
Specifically, to mitigate the issue of inaccurate speaker assignment, we first discard segments with a similarity lower than 0.5 to the cluster center for each speaker cluster. Additionally, to prevent cases where multiple people appear in a single cluster, we further discard clusters with an average similarity to the cluster center lower than 0.55 and a maximum similarity lower than 0.6.
To eliminate low-quality speech segments, we use the DNSMOS metric to evaluate each speech segment and discard those with an "OVRL" score lower than 2.4.

\vspace{-0.3cm}
\subsection{Speech recognition}
Following the quality filtering process, we employ a conformer-based neural transducer ASR system trained using approximately 60k hours of multi-lingual speech data to obtain the text transcriptions for each speech segment. The ASR system used in this study achieved WER of 4.51\%, 6.90\%, and 6.72\% on the WenetSpeech ``dev'', ``testnet'', and ``test meeting'' sets, respectively.




\vspace{-0.3cm}
\section{Experiments}
\label{sec: exp}
In order to analyze the performance of the proposed AutoPrep framework, we employ it to process the open-sourced WenetSpeech dataset and a self-collected in-the-wild dataset. To further evaluate the effectiveness of AutoPrep, we utilize the processed data in downstream TTS and speaker embedding extraction tasks.

\vspace{-0.3cm}
\subsection{Dataset}
\textbf{WenetSpeech}: WenetSpeech~\cite{zhang2022wenetspeech} is a widely used open-source ASR corpus, that comprises over 10,000 hours of Mandarin 16kHz speech data from diverse sources such as YouTube and Podcasts. Being derived from real-world data, WenetSpeech covers an extensive variety of acoustic conditions and includes a substantial number of speakers,  making it highly suitable for the application scenarios of AutoPrep.

\noindent \textbf{AutoPrepWild}: The AutoPrepWild corpus is a collection of in-the-wild speech data that we gathered from publicly available podcasts, video recordings, and audiobooks, without segmentation, speaker labels, or text transcriptions. The original dataset consists of 680 unprocessed long audio recordings, with a total duration of approximately 498 hours. Unlike the WenetSpeech dataset, the sample rate of the AutoPrepWild corpus is either 24kHz or 44.1kHz.

\begin{table}[htb]
    \centering
    \caption{Preprocessing results of AutoPrep framework on In-the-wild data and WenetSpeech dataset.}
    \scalebox{0.85}{
    \begin{tabular}{l|cc|cc|cccccccccc}
    \toprule
        Processing &Dur  & nSpk & DNSMOS &PDNSMOS & WER   \\
        \hline
        LJspeech         & 24h     & 1     & 3.30$\pm0.17$  & 4.17$\pm0.37$ &NA\\
        Aishell-3 \cite{shi2020aishell}       & 85h     &  218  & 3.15$\pm0.17$  & 3.83$\pm0.39$ &NA\\
        LibriTTS \cite{libritts}        & 586h    & 2456  & 3.25$\pm0.19$  & 3.97$\pm0.46$ &NA\\
        Studio-recorded & 156h        & 152  & 3.42$\pm0.12$   & 4.32$\pm0.28$ &NA\\
        \hline
        \hline
        AutoPrepWild         & 498h   & -  &2.05$\pm0.63$ &2.58$\pm0.61$ & 4.36 \\
        \cline{4-6}
         \ + SE      & 498h   & -  &\multirow{3}{*}{3.03$\pm0.40$} &\multirow{3}{*}{3.66$\pm0.68$} & \multirow{4}{*}{NA} \\
         \ + segmentation   & 240h   & -  & & &  \\
         \ + clustering     & 240h   & 372  & & &  \\
         \cline{4-5}
         \ + TSE            & 240h  & 372  &3.23$\pm0.22$ &4.06$\pm0.42$ & \\
         \cline{6-6}
         \ + Quality filter & 39h  & 48  &\textbf{3.24$\pm0.21$} &\textbf{4.07$\pm0.41$} & 4.01  \\
        \hline
        \hline
        WenetSpeech         & 10kh  & -  & 2.43$\pm0.55$  &2.76$\pm0.67$ & \multirow{5}{*}{NA}\\
        \cline{4-5}
         \ + SE      & 10kh  & -  & \multirow{2}{*}{2.84$\pm0.45$}  & \multirow{2}{*}{3.28$\pm0.76$} & \\
         \ + clustering     & 10kh   & 236.8k  & & & \\
        \cline{4-5}
         \ + TSE            & 10kh   & 236.8k  & 2.93$\pm0.38$ & 3.42$\pm0.67$ &  \\
         \ + Quality filter & 2840h & 60.5k  &\textbf{3.02$\pm0.27$} & \textbf{3.56$\pm0.56$} &    \\
    \bottomrule
    \end{tabular}
    }
    \label{tab:AutoPrepResults}
\end{table}

\begin{table*}[htb]
    \centering
    \caption{Results on DurIAN text-to-speech synthesis models trained with processed speech data generated by AutoPrep. Note that ``sub'' means a subset from the original dataset,  ``GT'' means groundtruth,  and ``NA'' means not applied.}
    \scalebox{0.9}{
    \begin{tabular}{c|ccc|cccc|ccccccc}
    \toprule
        Dataset & Dur &nUtt &nSpk  & Seg & Spk & SE & Trans   &  SMOS & MOS & SS \\
        \hline
        \multirow{2}{*}{WenetSpeech-P(sub)}     &342h & 460k & 2.0k & GT & AutoPrep  & NA               & GT   & 3.12$\pm0.15$ &2.90$\pm0.10$   & 0.55              \\
                                                &342h  & 460k & 2.0k & GT & AutoPrep  & AutoPrep        & GT   & 3.59$\pm0.13$ &3.10$\pm0.09$   & 0.60                \\
        \cline{1-11}
        WenetSpeech-P                           & 2.84kh  &3.65M & 60.5k & GT & AutoPrep  & AutoPrep    & GT   & \textbf{3.60$\pm0.12$} &\textbf{3.23$\pm0.10$}   & \textbf{0.68}            \\
        \hline
        \hline
        AutoPrepWild (sub) &4.21h &3291 &4 & AutoPrep & AutoPrep & AutoPrep & AutoPrep                         & \textbf{3.80$\pm0.18$}          &\textbf{3.52$\pm0.08$}    & 0.64 \\ 
                             
    \bottomrule
    \end{tabular}
    }
    \label{tab:my_label}
\end{table*}

\vspace{-0.3cm}
\subsection{Processing results of AutoPrep}
Table~\ref{tab:AutoPrepResults} demonstrates the processing results of each step in the proposed AutoPrep framework for both the WenetSpeech and AutoPrepWild datasets. 
Note that the DNSMOS and PDNSMOS scores in Table~\ref{tab:AutoPrepResults} are computed by randomly sampling 10,000 utterances from each corpus for testing due to the computational constraints. 
The results of the AutoPrepWild dataset reveal following observations:
1) Based on the DNSMOS and PDNSMOS metrics, the use of the BSRNN SE model significantly improves the audio quality of the original data.
2) The VAD-based segmentation discards approximately half of the data in the original audio.
3) Speaker clustering identifies 372 distinct speakers, but after quality filtering, only 48 speakers are retained. This suggests that the speaker label selection in AutoPrep exhibits strict criteria, prioritizing precision over recall.
4) The TSE model can further enhance the overall speech quality, especially in terms of PDNSMOS.
6) The quality filter does not provide significant improvement in the two MOS scores, suggesting that the speech quality for most of the speech segments in AutoPrepWild is acceptable after the SE and TSE operations.
7) On a small human-annotated test dataset consisting of 3026 segments from the AutoPrepWild data, the adopted ASR system achieves a WER of 4.01\%, indicating that the transcription accuracy of AutoPrep may still require improvement to reduce gap from human annotators. However, the auto-transcription can still serve as a useful reference for human annotators.  
8) With AutoPrep, the processed AutoPrepWild speech demonstrates comparable or even better speech quality than widely used open-source TTS datasets, such as LJspeech, Aishell-3, and LibriTTS. However, there still remains a quality gap when compared to studio-recorded high-quality speech data.

Similar trends can also be observed in the WenetSpeech results. However, we can see that the quality filter operation noticeably improves the speech quality of WenetSpeech, indicating that there are still challenging acoustic conditions that cannot be addressed by the current SE and TSE models. Furthermore, we can observe that even after discarding 71.6\% of the data, the speech quality of the AutoPrep-processed data still lags behind that of open-source TTS datasets. This suggests that using WenetSpeech data alone to train TTS models may present significant challenges in terms of audio quality. Note that in this experiment, we directly use the original segmentation and transcription provided in the WenetSpeech corpus. 

\begin{figure}
    \centering
    \includegraphics[width=1\linewidth]{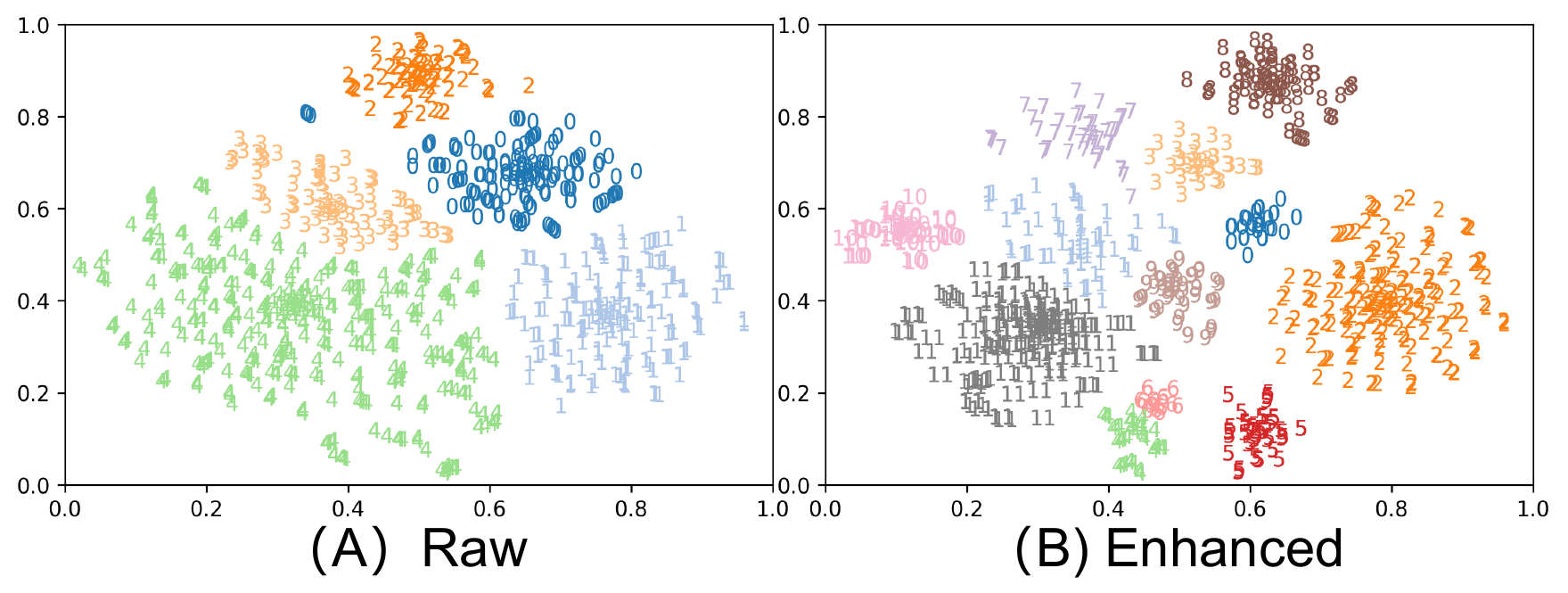}
    \caption{T-SNE Visualization of speaker clustering for utterance "Y0000007481\_J1rSmaVd4aE" in WenetSpeech.}
    \label{fig:cluster}
\end{figure}

\vspace{-0.3cm}
\subsection{Importance of speech enhancement}
In addition to noise removal and improving speech quality, speech enhancement, as the first step in the AutoPrep framework, also plays a crucial role in subsequent speaker clustering and ASR processes.
Figure~\ref{fig:cluster} illustrates the T-SNE visualization results of speaker clustering for a long audio from the WenetSpeech dataset (Y0000007481\_J1rSmaVd4aE). It is evident that for the original unprocessed audio, speaker clustering identifies five speakers, while the denoised audio results in 12 speakers. Upon manual verification, we discovered that the actual number of speakers in this audio exceeds 12, and in the clustering results of the non-denoised audio, category 4 encompasses more than one speaker. This phenomenon has been consistently observed in our experiments, leading us to conclude that noise affects the speaker embedding model's discriminative ability, and using denoised speech for clustering is essential when we prioritize accurate speaker label identification. 
Moreover, as shown in Table~\ref{tab:AutoPrepResults}, the SE can also improve ASR accuracy.

\vspace{-0.3cm}
\subsection{Evaluation on text to speech synthesis}
To further validate the effectiveness of the AutoPrep framework, we use the data processed by AutoPrep to train a multi-speaker TTS model based on the DurIAN TTS model. Training a multi-speaker TTS system requires speaker labels, audio, and text information, enabling a comprehensive evaluation of AutoPrep's processing outcomes.
Based on the findings in Table ~\ref{tab:my_label}, the following can be observed:
1) With the data processed by AutoPrep, we can successfully train a multi-speaker TTS system. The in-domain speaker similarity, calculated using Wespeaker embeddings between the synthesized and reference speech from the same in-domain speaker, is above 0.5, demonstrating the accuracy of the generated speaker labels from AutoPrep.
2) Incorporating AutoPrep's speech enhancement significantly improves the mean opinion score (MOS) and speaker similarity MOS (SMOS) score, which again demonstrates the effectiveness speech quality improvement ability of the AutoPrep framework. 
Note that the MOS test is conducted by 15 annotators using 42 synthesized speech samples from 7 different in-domain speakers for the processed WenetSpeech (WenetSpeech-P) experiment, and 16 synthesized speech samples from 4 in-domain speakers for the processed AutoPrepWild experiment. 
The model using Wenetspeech data is trained from scratch. And the model using AutoPrepWild data is finetuned from the pretrained TTS model.

\vspace{-0.3cm}
\subsection{Evaluation on speaker embedding extraction}
This section demonstrates that using the processed WenetSpeech data from AutoPrep to train a speaker embedding model based on DINO~\cite{caron2021emerging} can also yield improved performance on the CNCeleb-Eval~\cite{li2022cn} dataset for the speaker verification task.
The training details of DINO systems adhere to the WeSpeaker recipe, with ECAPA-TDNN~\cite{desplanques2020ecapa} employed as the backbone architecture. As shown in Table~\ref{tab:sv_cnceleb}, the DINO system trained with WenetSpeech-P outperforms the counterpart on both EER and minDCF.

\begin{table}[!htb]
    \centering
    \caption{Results of DINO-based speaker verification systems on CNCeleb-Eval.}
    \begin{tabular}{c|cc}
    \toprule
        Dataset &   EER(\%) & MinDCF  \\
        \hline 
        WenetSpeech   & 15.40 & 0.605 \\
        WenetSpeech-P & \textbf{15.03} & \textbf{0.560} \\                      
    \bottomrule
    \end{tabular}
    \label{tab:sv_cnceleb}
\end{table}

\vspace{-0.3cm}
\section{Conclusion}
This paper proposed the AutoPrep framework to automatically annotate in-the-wild speech data and improve speech quality. 
Experiments conducted on WenetSpeech and AutoPrepWild datasets demonstrate that AutoPrep can generate high-quality speech data and corresponding annotations for downstream task as TTS.

\vfill\pagebreak



\footnotesize{
\bibliographystyle{IEEEbib}
\bibliography{strings,refs}
}
\end{document}